\documentclass[a4paper,11pt, ]{article}
\pdfoutput=1

\usepackage{jcappub} 

\usepackage{subcaption}

\usepackage[T1]{fontenc} 
\usepackage{amsmath}
\DeclareMathOperator{\arcsec}{arcsec}
\newcommand{\angstrom}{\textup{\AA}}
\usepackage[usenames, dvipsnames]{xcolor}
\usepackage{hyperref}
\usepackage[export]{adjustbox}
\usepackage{multirow}
\usepackage{tikz}
\usetikzlibrary{shapes.geometric, arrows}
\usepackage{amsmath}
\usepackage{calc}      
\usepackage{setspace} 
\usepackage{fixltx2e}
\usetikzlibrary{shapes,arrows}
\usetikzlibrary{positioning}

\title{Spectro-Imaging Forward Model of Red and Blue Galaxies}

\author[a]{Martina Fagioli, \footnote{Corresponding author.}}
\author[a]{Luca Tortorelli,}
\author[a]{J\"org Herbel,}
\author[a]{Dominik Z\"urcher,}
\author[a]{Alexandre Refregier,}
\author[a, b]{and Adam Amara}

\affiliation[a]{Institute for Particle Physics and Astrophysics, ETH Z{\"u}rich, 8093 Z{\"u}rich, Switzerland}
\affiliation[b]{Institute of Cosmology, University of Portsmouth, Burnaby Road, Portsmouth PO1 3FX, UK}

\emailAdd{martina.fagioli@phys.ethz.ch}
\emailAdd{torluca@phys.ethz.ch}
\emailAdd{joerg.herbel@phys.ethz.ch}
\emailAdd{dominikz@phys.ethz.ch}
\emailAdd{alexandre.refregier@phys.ethz.ch}
\emailAdd{adam.amara@port.ac.uk}

\abstract{For the next generation of spectroscopic galaxy surveys, it is important to forecast their performances and to accurately interpret their large data sets. For this purpose, it is necessary to consistently simulate different populations of galaxies, in particular Emission Line Galaxies (ELGs), less used in the past for cosmological purposes. In this work, we further the forward modeling approach presented in Fagioli et al. 2018, by extending the spectra simulator \texttt{U\textsc{spec}} to model galaxies of different kinds with improved parameters from Tortorelli et al. 2020. Furthermore, we improve the modeling of the selection function by using the image simulator \texttt{U\textsc{fig}}. We apply this to the Sloan Digital Sky Survey (SDSS), and simulate $\sim157,000$ multi-band images. We pre-process and analyse them to apply cuts for target selection, and finally simulate SDSS/BOSS DR14 galaxy spectra. We compute photometric, astrometric and spectroscopic properties for red and blue,  real and simulated galaxies, finding very good agreement. We compare the statistical properties of the samples by decomposing them with Principal Component Analysis (PCA). We find very good agreement for red galaxies and a good, but less pronounced one, for blue galaxies, as expected given the known difficulty of simulating those. Finally, we derive stellar population properties, mass-to-light ratios, ages and metallicities, for all samples, finding again very good agreement. This shows how this method can be used not only to forecast cosmology surveys, but it is also able to provide insights into studies of galaxy formation and evolution.}

\begin{document}
\maketitle
\flushbottom

\section{Introduction}\label{intro}

The current and vastly accepted description of the evolution of our Universe includes the idea that it is in a phase of accelerated expansion. A number of observables probe this scenario, including the Hubble parameter $H(z)$ and the angular diameter distance $D_{A}(z)$. These probes are linked to the (baryonic and not) matter content and properties of our Universe \citep{planck2018, eisenstein2005}. 

Any dynamical study, including that of the Universe as a whole, requires a working definition of a distance scale. What is used in cosmology are the so called standard rulers \citep{seo2013}, whose evolution with cosmic time and size are known. A very good candidate for that are the Baryon Acoustic Oscillations (BAOs) \citep{eisenstein2005}. BAOs encode the width of primordial density fluctuations, which have been propagating in the form of acoustic waves. With the help of General Relativity, these can be expressed in the form of $H(z)$ and $D_{A}(z)$. Also, because of hydrodynamics, BAOs result in over-densities of baryons in shells at the sound horizon scale around an initial baryon over-density. These over-densities are the locations where galaxies formed, inside dark matter halos, through gravitational collapse \citep{white1978, blumenthal1984, davis1985}.

This is the reason why understanding how galaxies are located in the Universe is of fundamental importance. In particular, galaxy 3D distribution needs to be explored. While figuring out the celestial coordinates of an object is nowadays not especially challenging, understanding how \textit{far} an object is from us, or, in other words, its redshift $z$, is not straightforward. Multi-band imaging can be used to estimate photometric redshifts, which however require precise calibration and may suffer from catastrophic outliers. The precise distance of an object from us can be accurately estimated by looking at how much the features on its spectrum are displaced with respect to their laboratory wavelength value. This is a measure of its spectroscopic redshift $z$. In order to do so on a large scale, spectroscopic redshift surveys are required. 

In the last decades, cosmology oriented surveys have collected data mainly on the population of the so-called Luminous Red Galaxies (LRGs) \citep{eisenstein2001}, see for example the Deep Extragalactic Evolutionary Probe (DEEP2) \citep{weiner2005}, the Very Large Telescope Deep Survey (VVDS) \citep{garilli2008} and the Baryon Oscillation Spectroscopic Survey (BOSS) \citep{schlegel2009}  within the Sloan Digital Sky Survey (SDSS) III \citep{eisenstein2011}. However, the newer-generation spectroscopic surveys like extended-BOSS (eBOSS), \citep{dawson2016}, the Dark Energy Spectroscopic Instrument (DESI) \citep{schlegel2015}, the 4m Multi-Object Spectroscopic Telescope (4MOST) \citep{dejong2012} and Subaru-Prime Focus Spectrograph (PFS) \citep{sugai2015} aim at collecting data of a larger volume, up to the $z\sim2$ population of Emission Lines, or blue, Galaxies (ELGs). These were precedently almost unexplored in cosmological studies due to technical limitations \citep{masters2014, favole2017}, as ELGs are faint targets with emission features lying in the near-infrared background-dominated region of the spectrum at those redshifts. The combination of the two samples of ELGs and LRGs, probing different regions and cosmic times, will provide the most complete 3D map of the Universe to date.

To forecast and properly analyze these future experiments, it is therefore necessary to be able to properly simulate a wider variety of galaxies than the in past. Also, to appropriately account for possible observational biases, the simulation of a spectroscopic survey must be preceded by that of its parent imaging survey, the survey used to select targets for spectroscopy. As spectroscopic surveys require long integration time, it is crucial to appropriately pre-select suitable targets. Reproducing this step avoids that differences in the comparison between data and simulations may be originated from, for example, different magnitude definitions or star-galaxy separations, rather than the simulation itself. Also, computing speed  is essential to simulate large amount of data, which is why we present here several software tools which have been developed in order to achieve this.

In this work, we aim at forward modeling a wide field imaging survey such that SDSS, and, consequently, its spectroscopic counterpart. The main idea of forward modeling \citep{refregier2014, herbel2017, bruderer2017} is to combine an (simple yet realistic) astrophysical model, e.g., the evolution of the luminosity functions of red and blue galaxies with cosmic time, with instrumental parameters, in order to generate realistic simulations. The model can be complexified if the need of more detailed physics appears clear. In forward modeling, data and simulations are analyzed in the same way and compared. The inputs of the simulations can be adjusted such that their agreement improves. This has been proven to be successful for wide-field imaging surveys \citep{bruderer2017, herbel2017}, a narrow-band imaging survey like PAUs (Physics of the Accelerating Universe Survey\footnote{\url{https://www.pausurvey.org}}) \citep{tortorelli2018}, and a spectroscopic survey like SDSS \citep{fagioli2018}, hereafter [F18]. After simulating SDSS imaging data, we performe cuts on both data and simulations to obtain suitable samples for red and blue (or bluer, as stated in the CMASS Sparse definition) galaxies. We then forward model SDSS spectra, and compare those to data.

It is important to note that successfully simulating galaxy spectra with this technique means also having a successful galaxy model as a starting point. Also, it means being able to recover basic galaxy properties, such that this method can be also useful in galaxy formation and evolution studies. In the course of this paper, we will show that these two goals have been fulfilled. The galaxy population model, consisting in two different redshift-dependent luminosity functions for red and blue galaxies, has been presented in \citep{herbel2017}, and fully updated in \citep{tortorelli2020}, by using wide-field CFHTLS imaging data \citep{boulade2000}. The model parameters derived in \citep{tortorelli2020} are those used to simulate images and spectra in this work. Its successful application to independent spectroscopic data is a further proof of the validity of the method. 

The interpretation of absorption (and emission) features tells us about the origin and evolution of galaxies, of their stellar and gas components \citep{burstein1984, worthey1994, worthey1997, trager1998, trager2000, trager2005, korn2005, poggianti2001, thomas2003a, thomas2003b, korn2005, schiavon2007, thomas2011, onodera2012, fagioli2016}. This is of fundamental importance for the field of galaxy evolution. In this work, we recovered basic stellar population properties (mass to light ratios, stellar ages, and stellar metallicities) through full spectral fitting \citep {ocvirk2006b, ocvirk2006a, koleva2009, cappellari2004} for simulated spectra and compared those to real data, finding very good agreement, proving the usefulness of this method for measuring a broad range of galaxy properties. 

This paper is constructed as follows. As in forward modeling a combination of input galaxy model and instrumental parameters is used, we review in  Section~\ref{model} the galaxy population model, from its foundations to its most recent update obtained through an Approximate Bayesian Computation (ABC) run. This final, updated model is what we used throughout the rest of this work. Then, in Section~\ref{data}, we describe our sample data, needed for both the imaging and the spectroscopic analysis of this work. In Section~\ref{simulation}, we describe our image and spectra simulations, respectively performed with the \texttt{Ultra Fast Image Generator} (\texttt{U\textsc{fig}}) \citep{berge2013}, and \texttt{U\textsc{spec}} [F18]. In Section~\ref{analysis}, we describe our analysis pipeline to obtain both photometric and spectroscopic measurements. In Section~\ref{results}, we comment on our findings and in Section~\ref{conclusions} we present our conclusions. 
\\
\\
Throughout this work, we use a standard $\Lambda$CDM cosmology with $\Omega_\mathrm{m}=0.3$, $\Omega_\Lambda= 0.7$ and $\mathrm{H}_0 = 70$ km $\mathrm{s}^{-1}\,\mathrm{Mpc}^{-1}$.

\section{Galaxy Population Model}\label{model}

\subsection{Model Description}

Galaxy spectra encode unique information about galaxy (light emitting) content, i.e., stars and gas. To better understand such information, we need to simulate realistic galaxy spectra. The underlying idea is to start from basic galaxy properties, which are used as inputs in the simulations, that together with instrumental parameters produce realistic simulated data, which can be images, or spectra. The model can be complexified to match observations, as outlined in Section~\ref{abc}. The model from which such properties are drawn is fully described in \citep{herbel2017}, and the model parameter values have been further updated in \citep{tortorelli2020}, as described in Section~\ref{abc}. We here describe the main ideas behind this model.

Galaxy redshifts and magnitudes (and spectral coefficients, as further described below) are drawn from galaxy luminosity functions $\phi$ (see e.g. \citep{johnston2011, beare2015}). The galaxy luminosity function describes the number of galaxies $N$ per comoving volume $V$ in Mpc and absolute magnitude $M$:

\begin{equation}
\phi(z,M)=\frac{dN}{dM\,dV}
\end{equation}

where $z$ denotes redshift. Galaxies are drawn from separate and redshift dependent luminosity functions for blue and red objects. The redshifts and absolute magnitudes are obtained by sampling from the corresponding luminosity function. Then Spectral Energy Distributions (SEDs) are modeled. SEDs of galaxies are defined as linear combinations of templates $T_{i}(\lambda)$ and coefficients $c_i$, where $T_{i}(\lambda)$ are the templates and $c_i$ come from a Dirichlet distribution \citep{dirichlet} of order five, as described in \citep{herbel2017}. This model made use of the NYU Value-Added Galaxy Catalog (NYU-VAGC\footnote{\url{http://sdss.physics.nyu.edu/vagc/}}) \citep{blanton2005}, using as templates the $kcorrect$ templates presented in \citep{blanton2007}. These are based on Bruzual $\&$ Charlot stellar templates \citep{bruzual2003}. The stellar templates span an age range from 1 Myr to 13.75 Gyrs, and total stellar metallicities of $Z = 0.0001, 0.0004, 0.004, 0.008, 0.02$ and $0.05$, where $Z=0.02$ is the value for solar metallicity. The templates have solar $\alpha$-to-iron ($[\alpha/$Fe]) ratios. 35 templates come from MAPPINGS-III \citep{kewley2001} models of emission from ionized gas, which is what allows us to simulate bluer galaxies with emission features. Different coefficients $c_i$ are used for blue and red galaxies, coming from different Dirinchlet distributions, and they are redshift evolving, as better described below. The coefficients $c_i$ and redshifts $z$ are then given as inputs for the spectra simulations. The luminosity function parameters and magnitudes are used to render galaxy images.

\subsection{Model calibration through Approximate Bayesian Computation}\label{abc}

\begin{table}
\centering
\begin{tabular}{l c c c}
\hline
\hline
 & & \textbf{Blue} & \textbf{Red} \\
\hline
\multirow{5}{*}{Luminosity Function Parameters} &
$\alpha$ & -1.3 & -0.5 \\
&$\mathrm{M^*_{B,slope}}$ &-0.417 & -0.610 \\
&$\mathrm{M^*_{B,intcpt}} - 5 \log{h_{70}} $ & -20.591 & -20.416\\
&$\mathrm{\phi^*_{amp}}$ / 10$^{-3}$ h$^3_{70}$ Mpc$^{-3}$ mag$^{-1}$ & 0.0063& 0.0141 \\
&$\mathrm{\phi^*_{exp}}$ & -0.264 & -2.232 \\
\hline
\multirow{3}{*}{Size Parameters}
&$\mathrm{r_{50,slope}^{phys}}$ & -0.243 & -0.243 \\
&$\mathrm{r_{50,intcpt}^{phys}}$ & 0.954 & 0.954 \\
&$\mathrm{\sigma_{phys}}$ & 0.568& 0.568 \\
\hline
\multirow{10}{*}{Spectral Coefficients}

&$\mathrm{a}_{1,0}$ & 2.079& 2.461\\ 
&$\mathrm{a}_{2,0}$ & 3.524& 2.358\\
&$\mathrm{a}_{3,0}$ & 1.917& 2.568\\
&$\mathrm{a}_{4,0}$ & 1.992& 2.268 \\
&$\mathrm{a}_{5,0}$ & 2.536& 2.402 \\
&$\mathrm{a}_{1,1}$ & 2.265& 2.410 \\ 
&$\mathrm{a}_{2,1}$ & 3.862& 2.340 \\
&$\mathrm{a}_{3,1}$ & 1.921& 2.200 \\
&$\mathrm{a}_{4,1}$ & 1.685& 2.540 \\
&$\mathrm{a}_{5,1}$ & 2.480& 2.464\\
\hline
\hline
\end{tabular}
\caption{Median values of the galaxy model parameters for red and blue galaxies. In this table, the first 5 entries are the numerical values for the luminosity function parameters of Equation~\ref{lumfan}. $\mathrm{r_{50,slope}^{phys}}$, $\mathrm{r_{50,intcpt}^{phys}}$ and $\mathrm{\sigma_{phys}}$  are size parameters also derived in \citep{tortorelli2020}. The spectral coefficients are $\mathrm{a}_{\mathrm{i},0}$, which describe the galaxy population at $z = 0$, and $\mathrm{a}_{\mathrm{i},1}$, at redshift $z = z_1 > 0$. The spectral coefficients come from a Dirichlet distribution of order $5$.}
\label{tab: values}
\end{table}

The galaxy population model described above is calibrated in \citep{tortorelli2020}. In that work, we use Approximate Bayesian Computation (ABC) \citep{akeret2015} to constrain the galaxy population model parameters by matching simulations and data in an iterative way. ABC is a technique to approximate a Bayesian posterior by restricting the prior space in an iterative way, for cases where no clear likelihood can be provided. The prior space is reduced by minimising one or more distance metrics, which are described in details in \citep{tortorelli2020}. The data used in that work are images coming from Canada-France-Hawaii Telescope Legacy Survey (CFHTLS) wide-field galaxy survey \citep{boulade2000}. Image simulations are performed with \texttt{U\textsc{fig}}. As explained above, the aim of that work is to calibrate the galaxy population model and consequently measure the luminosity function of red and blue galaxies. 

The luminosity functions of red and blue galaxies can be expressed as 

\begin{equation}\label{lumfan}
\Phi (\mathrm{M,z}) \mathrm{dM}\,=\,\frac{2}{5} \ln{(10)} \phi^*(z) 10^{\frac{2}{5} (\mathrm{M^*(z) - M}) (\alpha(z) + 1)} \mathrm{e}^{\left[ -10^{\frac{2}{5} (\mathrm{M(z)^* - M})} \right]} \mathrm{dM}
\end{equation}

where M is the absolute magnitude of a galaxy, $z$ its redshift, $\phi*$ the normalization of the Schechter function \citep{schechter1976}, M$^*$ defines where the luminosity function transitions from a power law to a decaying exponential function, and $\alpha$ sets the faint end slope. The model parameters obtained with this run are listed in Table~\ref{tab: values}, together with spectral coefficients and size parameters. The spectral coefficients evolve as 

\begin{equation}
\mathrm{ a_i \left( z \right) = \left( a_{i,0} \right)^{1 - z / z_1} \times  \left( a_{i,1} \right)^{z/z_1}}
\end{equation}

where $\mathrm{a}_{\mathrm{i},0}$ describes the galaxy population at $z = 0$, while $\mathrm{a}_{\mathrm{i},1}$ at redshift $z = z_1 > 0$. The spectral coefficients come from a Dirichlet distribution of order $5$ and their values, which can be also found in Table~\ref{tab: values}, are used to generate spectra in this work.

\section{Data}\label{data}

\subsection{Images}

\subsubsection{Image Retrieval}\label{subsec: retrieval} 

We use data from SDSS (Sloan Digital Sky Survey) DR14 (server path: \url{http://dr12.sdss.org/sas/dr14/eboss/photoObj/frames/301/}). Images and spectra were obtained at the 2.5m telescope of the Apache Point Observatory (APO) in Sunspot, New Mexico \citep{gunn2006}. Each SDSS image can be uniquely identified by a sequence of three number, namely run, camera column (or "camcol"), and field. All information about SDSS imaging data can be found at \url{http://www.sdss3.org/dr10/imaging/imaging_basics.php}. We randomly select camcol, field and run to construct the path to the images. We download 157,000 images, all of them in five bands  (\textit{ugriz}). Alongside with each image, we download a star catalog from GAIA DR2 \citep{gaia2016, gaia2018} through the VizieR database of astronomical catalogues \citep{ochsenbein2000}, within a radius of $0.3$ degrees around the central pixel coordinates in each field. The use of this star catalog is explained in Section~\ref{par: simparest}. 

\subsubsection{Image Processing} 

In order for the images to be usable for forced photometry and spectra retrieval, some steps are necessary to pre-process them. Also, we need to compute some instrument-specific parameters to be able to properly simulate SDSS-like images. In the following paragraphs, we explain this procedure in details.

\paragraph{Alignment}

SDSS images are downloaded in 5 bands ($ugriz$). The same field in different bands shows a dithering of $\sim5-15$ pixels. This makes the computation of magnitudes with \texttt{SE\textsc{xtractor}} forced photometry incorrect. Forced photometry uses a reference image for the source coordinates, and then draws apertures and computes fluxes according to this reference. Therefore, a misalignment would result in a wrong magnitude computation. 

To align the field with pixel precision in the different bands we use \texttt{SW\textsc{arp}} \citep{bertin2002}. \texttt{SW\textsc{arp}} is a C-based software that resamples and co-adds together FITS images using an astrometric projection chosen by the user among a list in the World Coordinate System (WCS) standards. We use the standard \texttt{param.swarp} given configuration file, except for those keywords which are survey specific.

\paragraph{Astrometry}
The alignment of SDSS images in five different bands using  \texttt{SW\textsc{arp}} introduces an error in their astrometry. To fix this issue, we run \texttt{SCAMP} \citep{bertin2006}. \texttt{SCAMP} reads \texttt{SE\textsc{xtractor}} \citep{bertin1996} catalogs and computes astrometric and photometric solutions for any arbitrary sequence of FITS images. For this purpose, we first run \texttt{SE\textsc{xtractor}} with \texttt{SCAMP} specific parameters. \texttt{SCAMP} computes the astrometric solution by using as a reference the GAIA catalog described in~\ref{subsec: retrieval}. Then we update the FITS headers of SDSS images with the new computed WCS specific keywords.

\paragraph{Simulation Parameters Estimation}\label{par: simparest}
\begin{itemize}
\item \textit{PSF: } As described in Section~\ref{subsec: retrieval}, each set of images in the five $ugriz$ bands has an associated star catalog coming from GAIA. We use the star catalog to estimate the seeing and beta ($\beta$) values of the circular Moffat function, which is assumed to describe the star 1D light profile in SDSS \citep{xin2018}. We match the GAIA catalog to each image to securely identify stars. We then select stars within the magnitude range $14 < \mathrm{mag \,g} < 16$, to avoid faint or saturated stars. We make cutouts around the matched stars in the SDSS images. We then fit a two-dimentional Moffat profile, and consider the fitted median Full Width Half Maximum (FWHM) and $\beta$ as the input values for our \texttt{U\textsc{fig}} simulations.

\item \textit{Background: } The background is directly estimated from SDSS images. We first run \texttt{SE\textsc{xtractor}} to obtain a segmentation image, i.e., an image that separates objects from the sky background. Taking values coming only from regions identified as sky background, we compute the mean $\mu$ and the standard deviation $\sigma$, applying a 4$\sigma$ clipping. We then use those as parameters to simulate the gaussian sky background in \texttt{U\textsc{fig}}.

\item \textit{Magnitude Zeropoint: } The magnitude zeropoint in SDSS is computed as $-2.5\,\log_{10}f$, where $f$ is the flux expressed in maggies (1 maggie is the flux density in Janskys divided by 3631). The $f$ value in nanomaggies ($10^{-9}$ maggies) is contained in the keyword `NMGY' in the header of the SDSS fits images. It has been found that the $u-$band value needs to be shifted by $0.04$ mag in order to match the standard AB magnitude system \citep{knapen2014}, which we do in our simulations.
\item \textit{Pixel Scale: } The pixel scale for SDSS images is $0.396\arcsec$/pixel.
\item \textit{Gain: } Gain values in SDSS are specific for each camcol and filter. The full list can be found at \url{https://data.sdss.org/datamodel/files/BOSS_PHOTOOBJ/frames/RERUN/RUN/CAMCOL/frame.html}.
\item \textit{Saturation: } The imaging data saturate at about 13, 14, 14, 14, 12 magnitudes for point sources\footnote{\url{https://classic.sdss.org/dr7/instruments/technicalPaper/index.html}}.
\item \textit{Exposure time and filters: } The five filters are observed in order \textit{r i u z g}, 71.7 seconds apart. The integration time in each filter is $\sim54$ seconds.
\end{itemize}

\subsection{Spectra}\label{specretriv}

Spectra are downloaded through the \texttt{astroquery} package \citep{ginsburg2019}, from SDSS DR14 spectroscopic data release. We use the RA and DEC coordinates stored in the \texttt{SE\textsc{xtractor}} variables (ALPHAWIN\_J2000 and DELTAWIN\_J2000) and query spectra within $1\arcsec$ searching radius. We only download spectra observed with the BOSS  (Baryon Oscillation Spectroscopic Survey) instrument \citep{schlegel2009, eisenstein2011}, and with $z>0$, so to avoid negative or nan values.

\section{Simulations}\label{simulation}
\subsection{Image Simulation}

\begin{figure}
\centering

	\includegraphics[width=155mm]{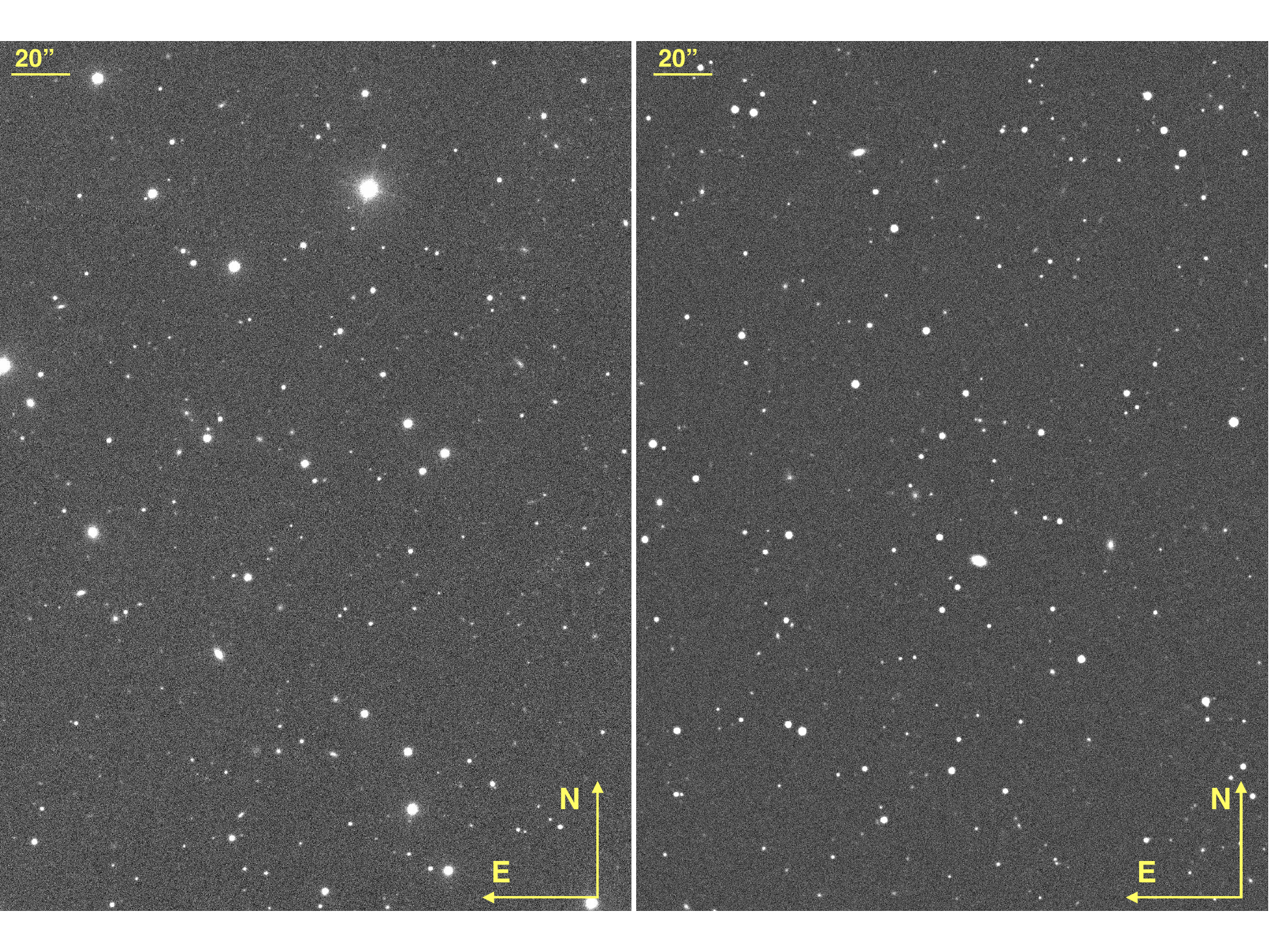}

    \caption{Comparison between a SDSS real (left panel) and simulated (right panel) $i-$band image. The real SDSS image has been preprocessed as described in Section~\ref{data}. The north and east directions are shown at the bottom of the panels with yellow arrows. The upper left thick shows the angular size on the image corresponding to $20\arcsec$.}
    \label{fig: image}
\end{figure}

We simulate SDSS imaging data with the \texttt{Ultra Fast Image Generator} (\texttt{U\textsc{fig}}) \citep{berge2013, bruderer2016, bonnett2016, leistedt2016, herbel2017, bruderer2017, tortorelli2018, kacprzak2019, tortorelli2020}. \texttt{U\textsc{fig}} simulates astronomical images by combining parameters coming from a galaxy population model (magnitudes, sizes and S\'ersic profiles) as described in Section~\ref{model}, and instrumental parameters coming from the survey we aim at simulating. \texttt{U\textsc{fig}} ability of simulating a variety of wide- (DES, COSMOS, CFHTLS) and narrow- (PAUS) band survey has been shown in \citep{berge2013, bruderer2016, bonnett2016, leistedt2016, herbel2017, bruderer2017, tortorelli2018, kacprzak2019, tortorelli2020}. This is the first time \texttt{U\textsc{fig}} is used to render SDSS images. We use the instrumental properties derived as described in Section~\ref{par: simparest}. Figure~\ref{fig: image} shows the comparison between a real image from SDSS (left hand side) and a \texttt{U\textsc{fig}} simulated image (right hand side). We run tests to assess the reliability of our image simulation, comparing the pixel and number counts for real and simulated images \citep{berge2013}, finding very good agreement. For further details on \texttt{U\textsc{fig}} and particularly on its speed, we refer the reader to \citep{berge2013}.

\subsection{Spectra Simulation}

We simulate galaxy spectra with our software \texttt{U\textsc{spec}}. \texttt{U\textsc{spec}} has been presented and described in details in [F18]. \texttt{U\textsc{spec}} usage is fully outlined in the flowchart in Figure 3 in [F18]. Briefly, \texttt{U\textsc{spec}} takes as inputs the galaxy model described in Section~\ref{model}, and also in \citep{herbel2017} and \citep{tortorelli2020}, and the instrumental setup (read-out noise, shot noise, transmission curve, exposure time) of a given telescope. As an update from the previous version of \texttt{U\textsc{spec}}, we now include the effects of Galactic dust, following the extinction curve for diffuse gas from \citep{odonnell1994} with $R_v=3.1$, and using the Galactic $E(B-V)$ values from the maps of \citep{schlegel1998}, in our spectra simulations. It is important to mention that one of the main noise contaminants in galaxy spectra is the spectrum coming from our own atmosphere. In [F18], we explain in details how we account for this effect, and in Figure 4 of [F18] we show the sky model which is also used in this work, and how we constructed it. In the end, \texttt{U\textsc{spec}} outputs redshifted, noisy galaxy spectra, for both blue and red galaxies. This differentiation derives from the different spectral coefficients coming by different luminosity functions for red and blue galaxies, drawn together with magnitudes and redshifts. These spectral coefficients are combined with $kcorrect$ spectral templates \citep{blanton2007} to produce galaxy model spectra.

\section{Analysis}\label{analysis}

\subsection{Imaging Measurements}\label{photmes}

\begin{figure}
\centering

	\includegraphics[width=135mm]{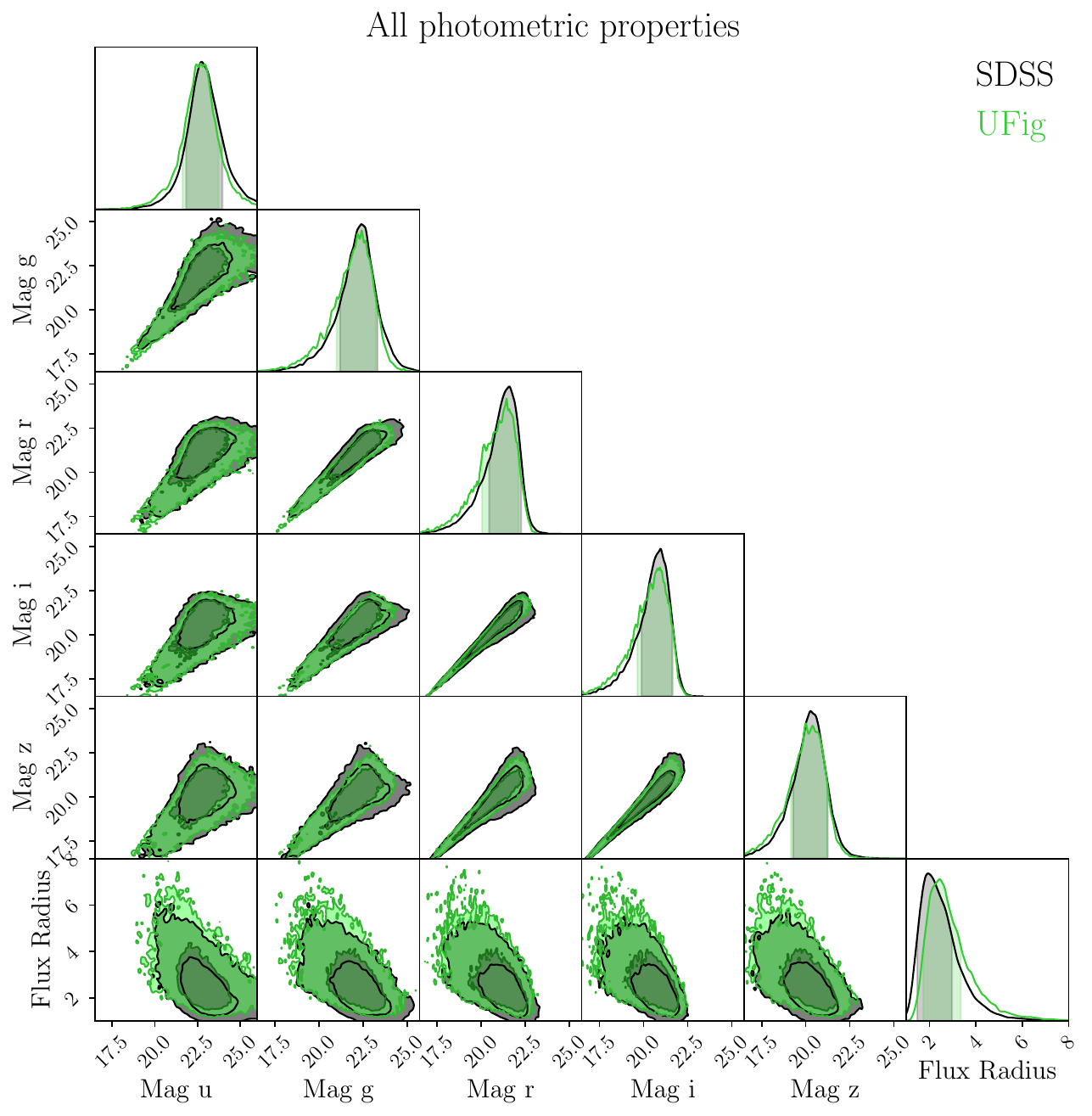}

    \caption{Comparison of galaxy imaging properties between real (black) and simulated (green) data. The 5 ($ugriz$) SDSS band are shown, together with the \texttt{SE\textsc{xtractor}} FLUX\_RADIUS parameter. The magnitude distributions are the $\mathrm{mag}_{\mathrm{corr}}$ defined in Equation~\ref{eq: corr} in Section~\ref{photmes}. The agreement between data and simulation is particularly noteworthy as in no stage of the galaxy model parameters calibration in \citep{tortorelli2020} SDSS images have been used.}
    
    \label{fig: magnitudes_tot}
\end{figure}
The large amount of data, and, accordingly, simulations, required for this work, made the use of the multiple cores of the Euler cluster\footnote{\url{https://scicomp.ethz.ch/wiki/Main\_Page}} necessary. The whole analysis and simulations described above take about 4 hours with 1,000 cores.

We perform imaging measurements on both simulated and real images with \texttt{SE\textsc{xtractor}}. On each pair of real and simulated SDSS fields (the real field from SDSS and its analog, i.e., a simulated image with instrumental parameters measured as described in Section~\ref{par: simparest}), we run \texttt{SE\textsc{xtractor}} in forced photometry mode in the five $ugriz$ bands. In order to do so, we construct a detection image as described in \citep{coe2006} and summarized in \citep{tortorelli2020}, and references therein. In brief, we stack the $ugriz$ images and normalize each of them by their background $\sigma$ computed as described in~\ref{par: simparest}. We then used the maximum seeing, magnitude zeropoint, saturation and gain within the five images. The use of the maximum seeing ensures we do not loose signal coming from the source. This is further guaranteed by the alignment of the images made through \texttt{SW\textsc{arp}}. We then measure magnitudes in each band by using the aperture fixed in the detection image, and correct those for the different PSF among bands as described in \citep{coe2006} and as shown below:

\begin{equation}\label{eq: corr}
\begin{array}{l}
u_{\mathrm{corr}} = u_{\mathrm{ISO}} - (d _{\mathrm{ISO}}- d_\mathrm{AUTO})\\
g_{\mathrm{corr}} = g_{\mathrm{ISO}} - (d _{\mathrm{ISO}}- d_\mathrm{AUTO})\\
r_{\mathrm{corr}} = r_{\mathrm{ISO}} - (d _{\mathrm{ISO}}- d_\mathrm{AUTO})\\
i_{\mathrm{corr}} = i_{\mathrm{ISO}} - (d _{\mathrm{ISO}}- d_\mathrm{AUTO})\\
z_{\mathrm{corr}} = z_{\mathrm{ISO}} - (d _{\mathrm{ISO}}- d_\mathrm{AUTO}),
\end{array}
\end{equation}

where $d _{\mathrm{ISO}}$ and $d_\mathrm{AUTO}$ are the \texttt{SE\textsc{xtractor}} MAG\_ISO and MAG\_AUTO measured on the detection image. We measure and store also FLUX\_RADIUS and coordinates RA and DEC for later usage. The same procedure is applied to data and simulations, as in the philosophy of forward modeling.

Figure~\ref{fig: magnitudes_tot} shows the agreement between our real (black) and simulated (green) $\mathrm{mag}_{\mathrm{corr}}$ and the \texttt{SE\textsc{xtractor}} FLUX\_RADIUS parameter. This agreement on SDSS imaging measurements is a further proof of the goodness of the luminosity function parameters computed in \citep{tortorelli2020}, as it shows that same parameters can be applied to different surveys than those used in the ABC run. It is noteworthy that in no stage of the ABC run SDSS was used to constrain the luminosity function parameters, and yet here it is shown that SDSS images can be properly simulated with those parameters.

\subsection{Selection Cuts}

\begin{figure}	
	\centering
	\begin{subfigure}[t]{0.4955\linewidth}
		\centering
		\includegraphics[width=1\linewidth, valign=t]{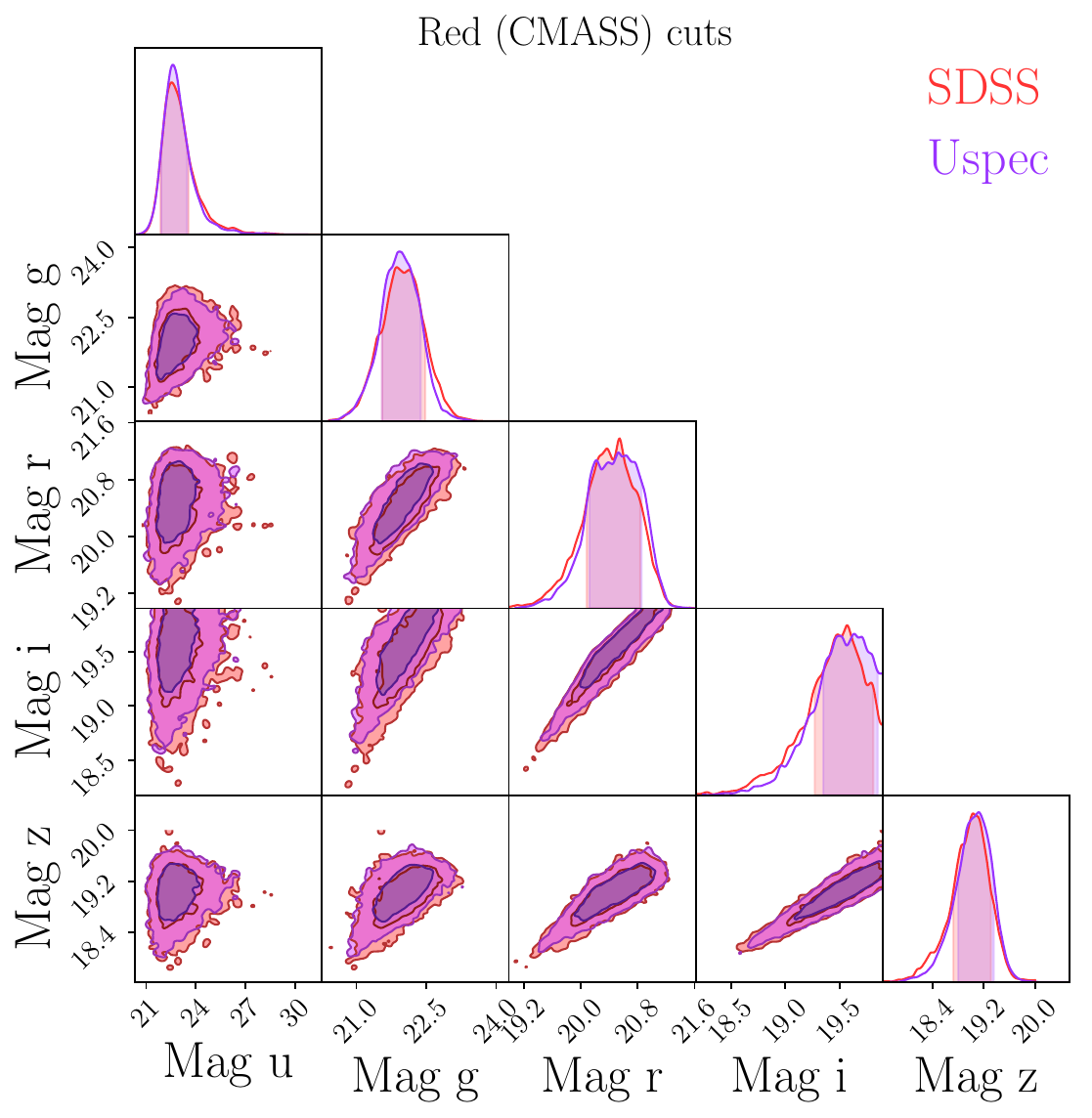}
	\end{subfigure}
	\begin{subfigure}[t]{0.4955\linewidth}
		\centering
		\includegraphics[width=1\linewidth,valign=t]{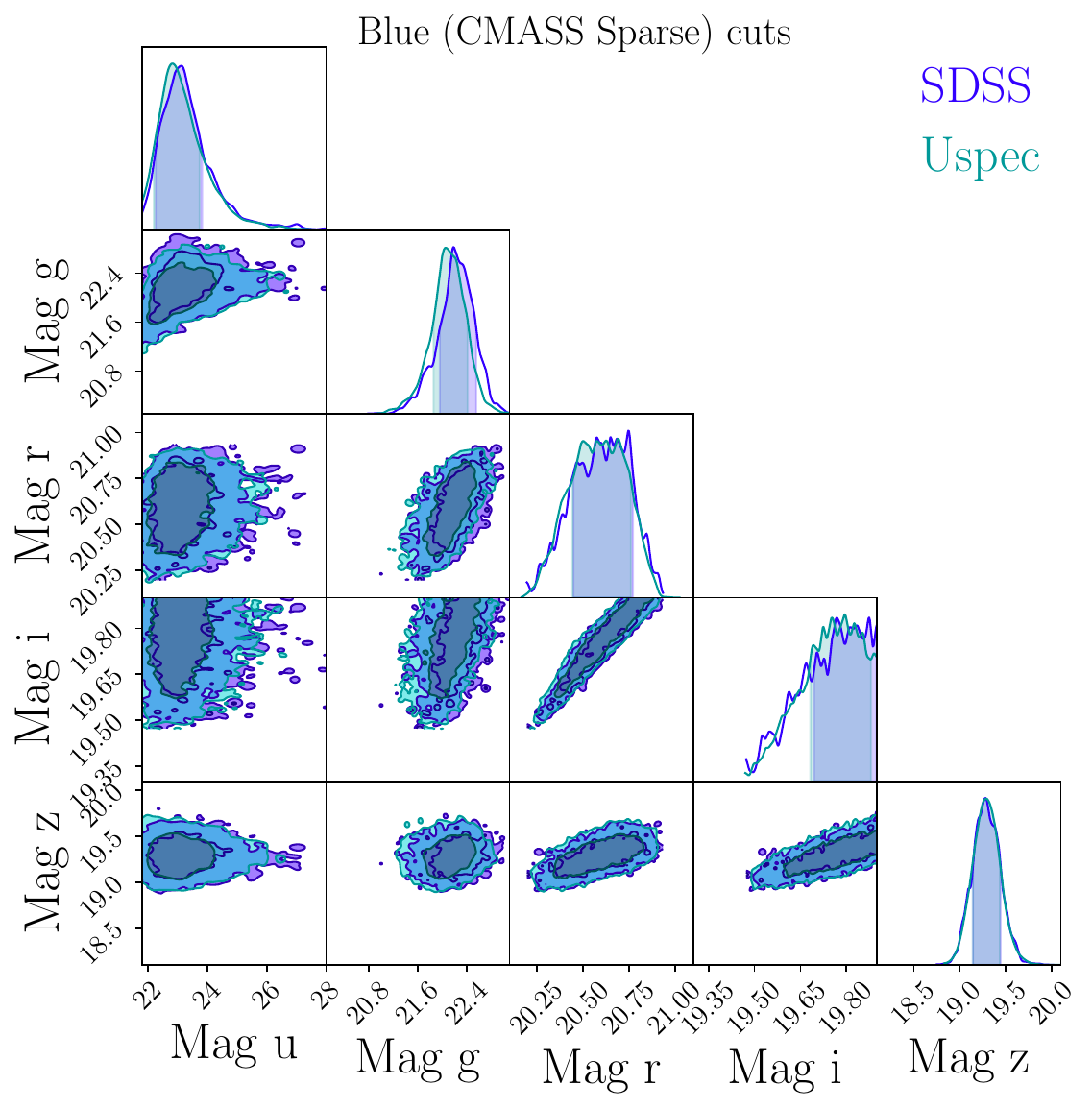}

	\end{subfigure}
	\caption{Comparison between photometric properties of real and simulated data for red (left panel) and blue (right panel) galaxies. Both panels show magnitudes of objects after applying the CMASS (for red galaxies) and CMASS Sparse (for blue galaxies) cuts. For all galaxies shown in this panel we retrieved the spectra which are later used in our spectroscopic analysis.}
	\label{fig: magnitudes_cuts}
\end{figure}

We emulate cuts as in the SDSS/BOSS CMASS\footnote{\url{http://www.sdss3.org/dr9/algorithms/boss_galaxy_ts.php}} sample \citep{dawson2013}. For a more extensive description of these cuts, please read [F18], and references therein. We call this sample the red galaxies sample throughout the rest of this paper. In addition, we emulated the CMASS Sparse sample cuts. We call this sample the blue galaxies sample throughout the rest of this paper, as this is defined to be a sample of fainter and bluer galaxies with respect to CMASS. CMASS Sparse has been designed to randomly select 1 in about 10 targets. \\
Our specific magnitude (i.e., our $\mathrm{mag}_{\mathrm{corr}}$ as described above), color, and star-galaxy separation cuts are listed below:

\paragraph{CMASS, or Red Galaxies}
\begin{itemize}
\item $17.5<\mathrm{mag}\,i<19.9$
\item $r-i<2$
\item $d_\perp>0.55$, where $d_\perp = (r-i) - (g-r) / 8$
\item $\mathrm{mag}\,i<19.86+1.6 \times (d_\perp-0.8)$
\item \texttt{SE\textsc{xtractor}} CLASS\_STAR < 0.1, for star-galaxy separation
\end{itemize}

\paragraph{CMASS Sparse, or Blue Galaxies}\label{par: cmasssparse}
\begin{itemize}
\item $17.5<\mathrm{mag}\,i<19.9$
\item $r-i<2$
\item $d_\perp>0.55$, where $d_\perp = (r-i) - (g-r) / 8$
\item $\mathrm{mag}\,i<20.14+1.6 \times (d_\perp-0.8)$
\item \texttt{SE\textsc{xtractor}} CLASS\_STAR < 0.1, for star-galaxy separation
\end{itemize}

The \texttt{SE\textsc{xtractor}} CLASS\_STAR < 0.1 has been used to ensure purity of the sample from stars or quasars. However, no appreciable differences can be seen when applying a  CLASS\_STAR cut of $<0.9$.

Figure~\ref{fig: magnitudes_cuts} shows the comparison between real and simulated magnitudes after applying the above cuts for red and blue galaxies. The two samples show a good agreement overall, with however a small shift towards fainter magnitudes for SDSS in the central $gri$ bands. A similar difference can be seen in Figure 7 in \citep{tortorelli2020}. This is likely to be an observational effect due to BOSS targeting that our forward modeling technique is not able to reproduce at the current stage, as we do not see any sign of this in the magnitude comparison for the whole sample (Figure~\ref{fig: magnitudes_tot}).

\begin{figure}	
	\centering
	\begin{subfigure}[t]{0.48\linewidth}
		\centering
		\includegraphics[width=1\linewidth, valign=t]{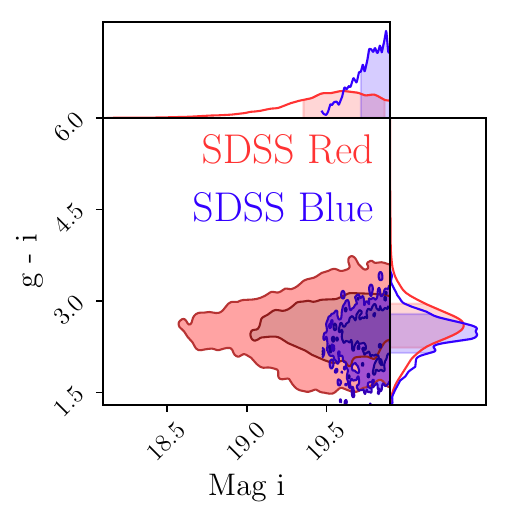}
	\end{subfigure}
	\begin{subfigure}[t]{0.48\linewidth}
		\centering
		\includegraphics[width=1\linewidth,valign=t]{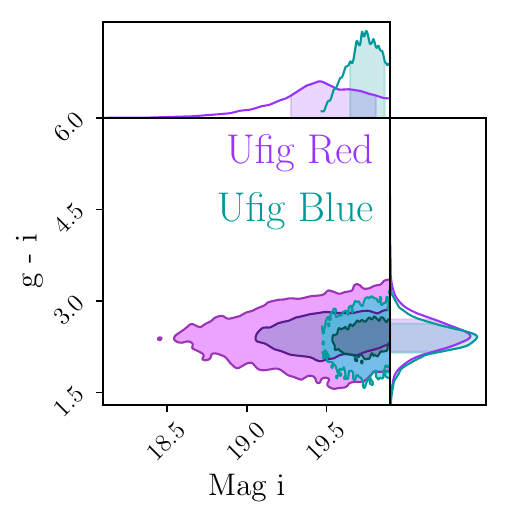}

	\end{subfigure}
	\caption{Color (g-i) - magnitude ($i-$band) diagram for real and simulated data. We used the $\mathrm{mag}_{\mathrm{corr}}$ described in Section~\ref{photmes}. On the left hand side, a comparison between the CMASS (red) and CMASS Sparse (blue) galaxies from the SDSS survey. CMASS Sparse is defined to target bluer and fainter objects than the parent sample CMASS. This is visible also in our contours. The same trends are visible in the simulated samples on the right hand side on the plot.}
	\label{fig: color_cuts}
\end{figure}

Figure~\ref{fig: color_cuts} shows the color-magnitude comparison between SDSS CMASS and CMASS Sparse. On the left hand panel, a comparison between the $g-i$ color and the magnitude in the $i-$band of SDSS red (CMASS) and SDSS blue (CMASS sparse) samples is shown. The same on the right hand side panel, but for simulations. The two figures show how both data and simulation have the same behaviour, with the CMASS Sparse-like sample occupying the region of fainter magnitudes of the CMASS sample. CMASS Sparse is indeed described as a sample of bluer and fainter galaxies, and this is expressed in the cuts listed in Section~\ref{par: cmasssparse}.

\subsection{Spectroscopic Measurements}
After galaxies are selected according to the cuts described above, galaxy spectra for both the blue and the red samples are downloaded, as described in Section~\ref{specretriv}. The spectra are then analysed and compared with two different techniques: by using Principal Components Analysis (PCA), and by looking at their stellar population properties computed with full spectral fitting. The results coming from these two analysis are shown in Figures~\ref{fig: pca_red},~\ref{fig: pca_blue},~\ref{fig: coefficients} and~\ref{fig: gal_prop}, and described in details in Section~\ref{results}. Here we describe the methodologies we use for these two kinds of analysis.

\subsubsection{Principal Components Analysis}\label{pca}

Our use of PCA in comparing data and simulation has been already described in details in [F18] and \citep{tortorelli2018}. To briefly summarize, applying PCA means in this case representing the spectra as a set of eigenspectra of lower dimension, as also shown in \citep{connolly1995}. We use Singular Value Decomposition to compute the eigenspectra and eigencoefficients, as sets of spectra can be described as

\begin{equation}
\label{specdata}
{f}(\lambda)=\sum_j \mathrm{a}_{j}\phi_j(\lambda)\,~~\mathrm{for \,data}
\end{equation}

\begin{equation}
\label{specsim}
{f'}(\lambda)=\sum_j \mathrm{b}_{j}\psi_j (\lambda)\,~~\mathrm{for \,simulations}
\end{equation}
where a$_{j}$, or b$_{j}$, are the eigencoefficients, and $\phi_j (\lambda)$, or $\psi_j (\lambda)$, are the eigenspectra for the data or the simulations, accordingly.

These are computed independently for data and simulations, and can be compared through the Mixing Matrix. We define the Mixing Matrix as

\begin{equation}
\mathrm{M}_{ij}=\int\phi_i(\lambda)\psi_j(\lambda)~d\lambda
\end{equation}
so that, if $\phi_i (\lambda)\equiv\psi_i (\lambda)$, $\mathrm{M}_{ij}$ reduces to

\begin{equation}
\mathrm{M}_{ij}=\delta_{ij}
\end{equation}
which means that, if real and simulated data were described by the same basis set, the mixing matrix would be the Identity Matrix.

\begin{figure}	
	\centering
	\begin{subfigure}[t]{0.43\linewidth}
		\centering
		\includegraphics[width=1\linewidth, valign=t]{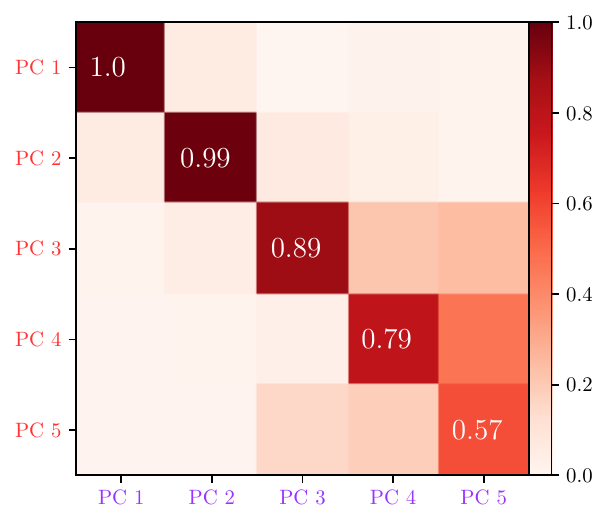}
	\end{subfigure}
	\begin{subfigure}[t]{0.43\linewidth}
		\centering
		\includegraphics[width=1\linewidth,valign=t]{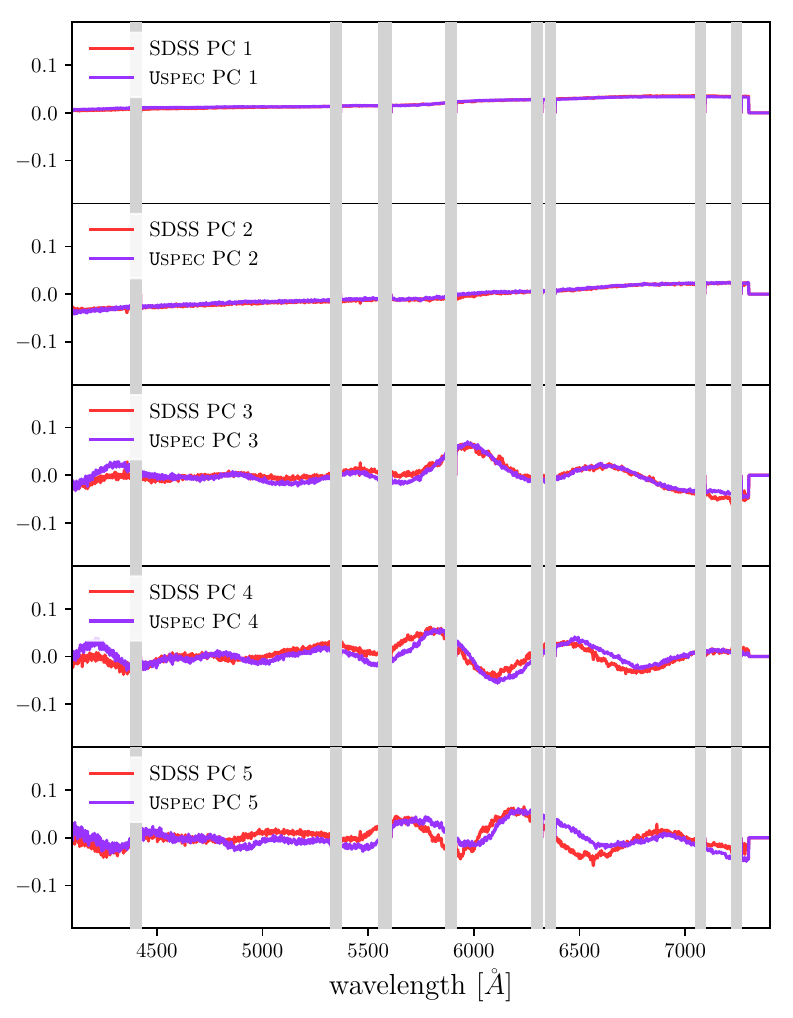}

	\end{subfigure}
	\caption{Mixing matrix and first 5 PCA components for red galaxies. The comparison between all first five principal components shows a very good agreement between data and simulations. This is quantified in the Mixing Matrix shown on the left hand side of the plot.}
	\label{fig: pca_red}
\end{figure}

For this analysis, we mask regions where strong sky lines are expected (4403 \AA, 5350 \AA, 5577 \AA, 5588 \AA, 5894.6 \AA, 6301.7 \AA, 6364.5 \AA , 7246.0 \AA, 7074 \AA\, with 20 \AA\, width, the grey stripes in Figures~\ref{fig: pca_red} and~\ref{fig: pca_blue}), and we exclude the $10\%$ brightest objects in the two samples, so that the outliers would not dominate the principal components. The data and simulations are analysed with matching their redshift distributions. Further details on this aspect can be found in Section~\ref{secspecprop}. This procedure is applied to both data and simulations.

\subsubsection{Full Spectral Fitting}\label{full}

We use full spectral fitting to compute stellar population parameters. Full spectral fitting is a technique which has been developed to compute stellar population parameters \citep{ocvirk2006b, ocvirk2006a, koleva2009, cappellari2004}, mostly stellar ages and stellar metallicities, but can also include stellar masses and gas properties. Fitting the whole spectrum at the same time constitutes an advancement on the older technique which consisted in fitting individual absorption features and their pseudo-continua \citep{burstein1984, worthey1994, worthey1997, trager1998, trager2000, trager2005, korn2005, poggianti2001, thomas2003a, thomas2003b, korn2005, schiavon2007, thomas2011, onodera2012, onodera2012, fagioli2016}. In this paper, we used the latest version of \texttt{Penalized Pixel-Fitting} (\texttt{pPXF}) \citep{cappellari2017} to perform full spectral fitting. We use templates from the MILES library \citep{vazdekis2016}, with a wide range of stellar ages and metallicities, and solar $[\alpha/$Fe] ratios. During the fit we use a 20th order multiplicative polynomial correction and no additive polynomials, as recommended in \citep{cappellari2017}. We do not fit the gas component simultaneously, to improve the speed of the fit. We derived stellar ages, metallicities and mass to $r-$band based light ratios as described in detail in Section~\ref{results}.

\section{Results}\label{results}

\subsection{Principal Components}\label{respca}

\begin{figure}	
	\centering
	\begin{subfigure}[t]{0.43\linewidth}
		\centering
		\includegraphics[width=1\linewidth, valign=t]{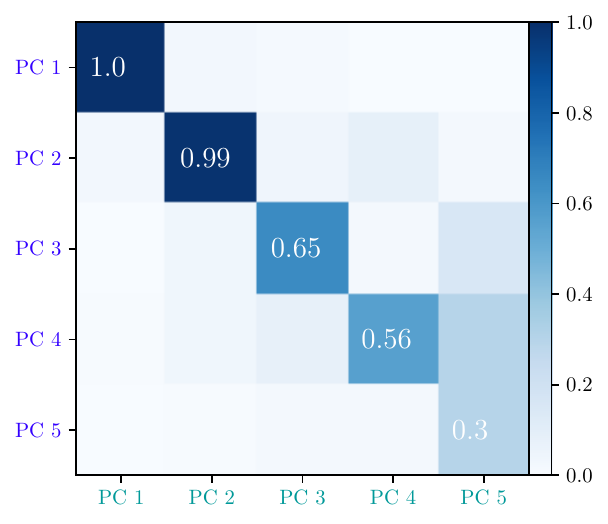}
	\end{subfigure}
	\begin{subfigure}[t]{0.43\linewidth}
		\centering
		\includegraphics[width=1\linewidth,valign=t]{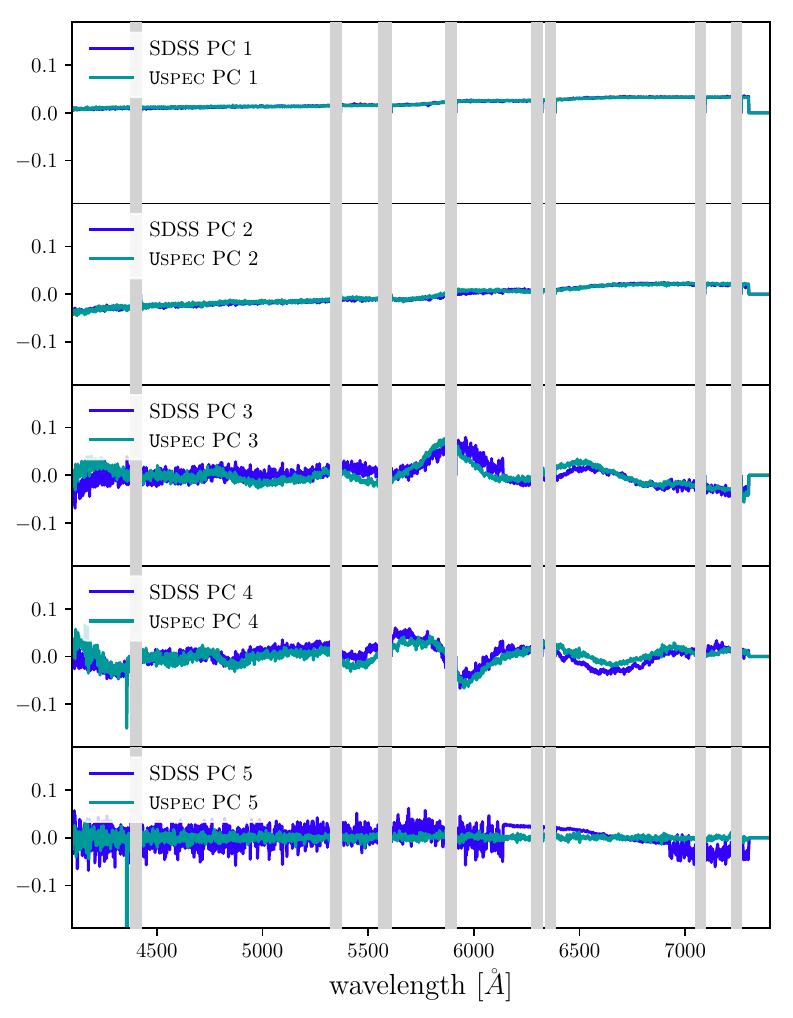}

	\end{subfigure}
	\caption{Mixing matrix and first 5 PCA components for blue galaxies. The comparison of the first two principal components shows excellent agreement. A less good agreement is shown for higher order principal components, as expected for blue galaxies, notoriously known as difficult to individually, rather than statistically, simulate. Higher order principal component are more sensitive to individual galaxy features. As in the previous Figure, the comparison between principal components is quantified in the Mixing Matrix shown on the left hand side of the plot.}
	\label{fig: pca_blue}
\end{figure}

In order to assess our ability to properly simulate galaxy spectra, we use PCA to quantify the agreement between our spectra simulation and SDSS data. The methodology has been described in details in [F18], and \citep{tortorelli2018}, and in Section~\ref{pca}. The first five eingenspectra, or principal components, we decomposed our galaxies into are shown in Figures~\ref{fig: pca_red} (red galaxies) and Figure~\ref{fig: pca_blue} (blue galaxies), in the right panels. In the left panels, we show the associated Mixing Matrices, as described in Section~\ref{pca}.

\begin{figure}	
	\centering
	\begin{subfigure}[t]{0.495\linewidth}
		\centering
		\includegraphics[width=1\linewidth, valign=t]{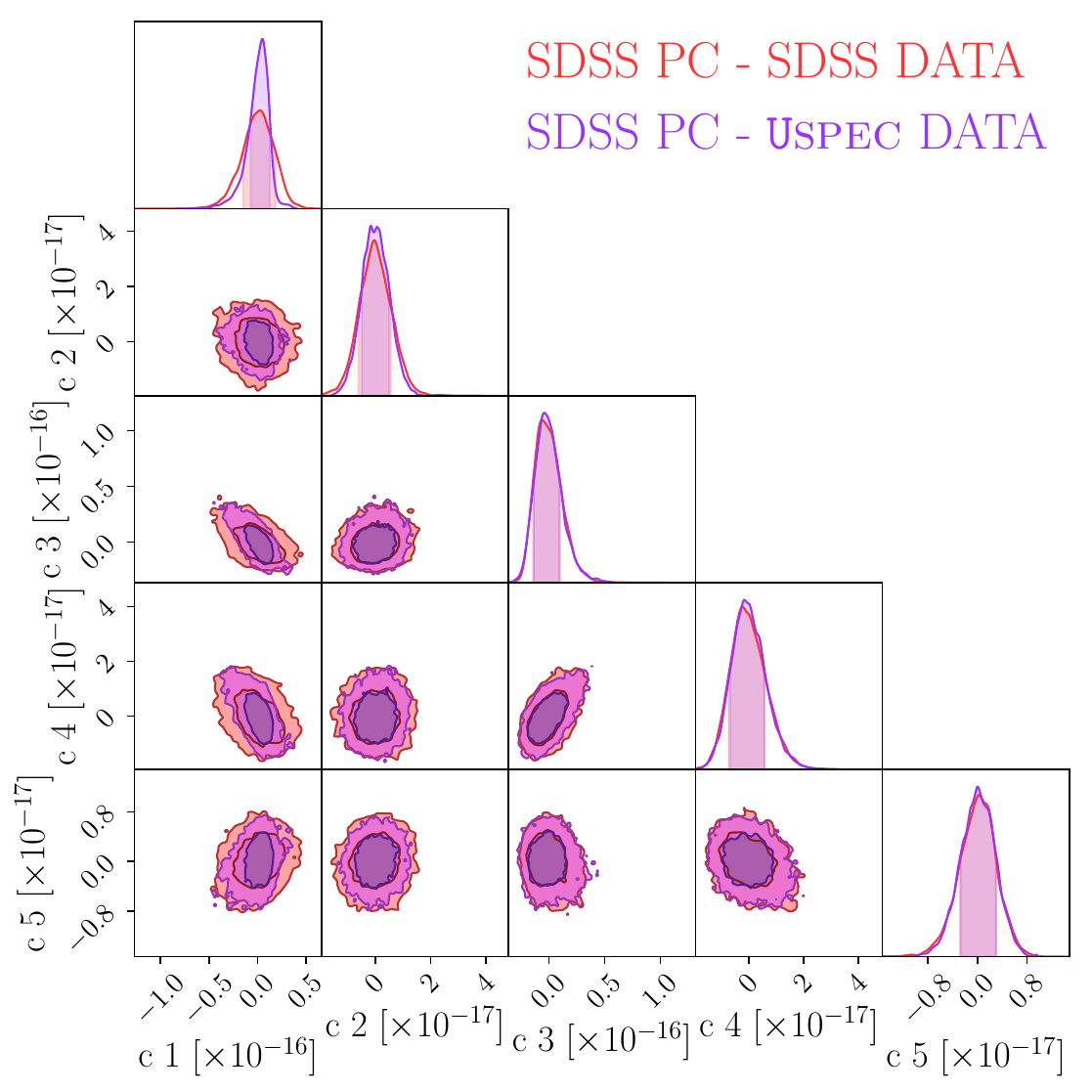}
	\end{subfigure}
	\begin{subfigure}[t]{0.495\linewidth}
		\centering
		\includegraphics[width=1\linewidth,valign=t]{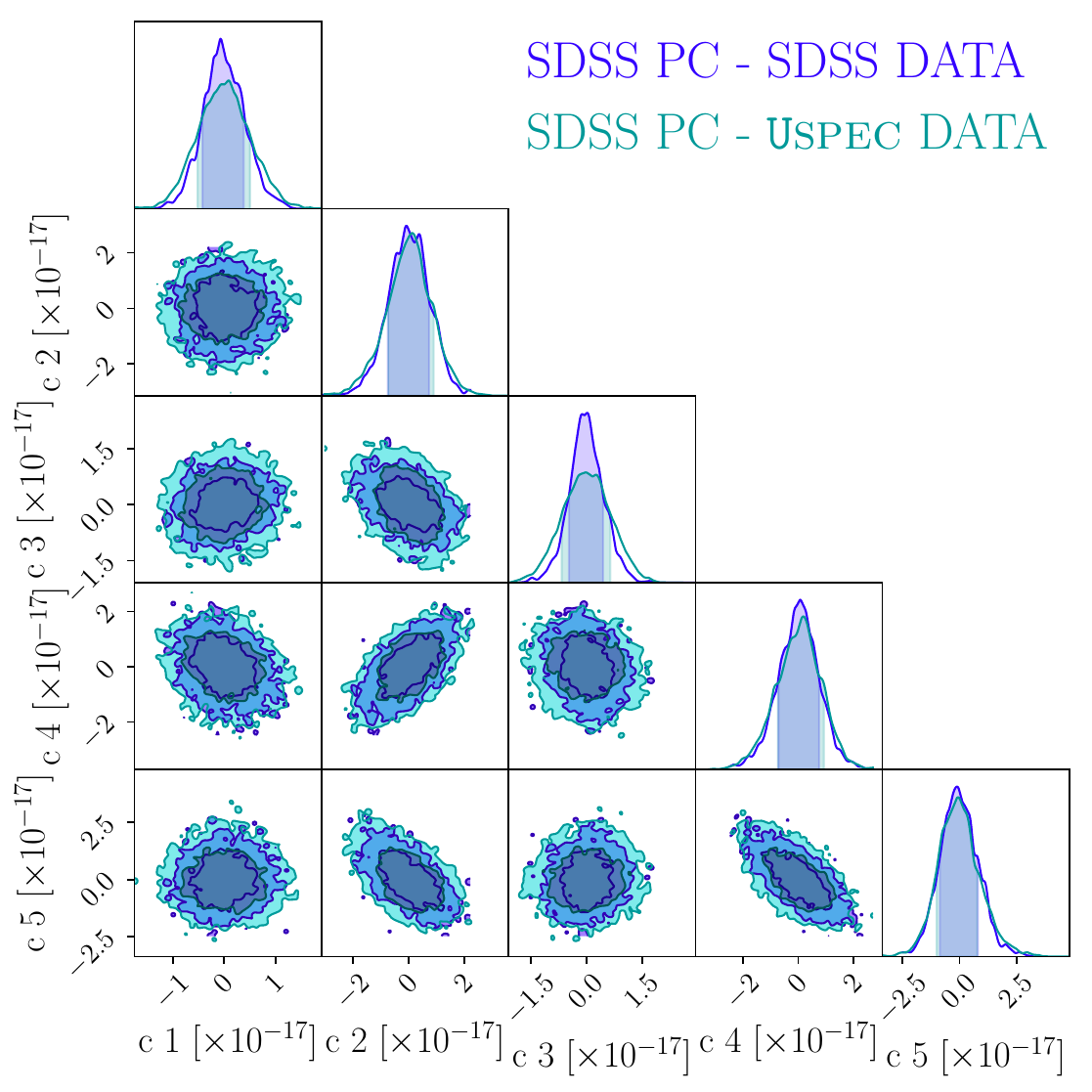}

	\end{subfigure}
	\caption{Eigencoefficients resulting from projecting real spectra from SDSS and simulated \texttt{U\textsc{spec}} spectra onto SDSS principal components. The distributions for all the first five principal components appear to be centered to each other and present the same width, for both red and blue galaxies. The eigencoefficients are expressed in the same units of flux ($10^{-17}\mathrm{ergs}\,s^{-1}\, \angstrom^{-1}\,\mathrm{cm}^{-2}$) as the real and simulated spectra.}
	\label{fig: coefficients}
\end{figure}

As seen in Figure~\ref{fig: pca_red}, red galaxies principal components show very good agreement. For the first component, which encodes most of the information coming from the galaxy population, we find very good agreement between data and simulations, showing that our galaxy population model, derived in \citep{tortorelli2020} and summarized in Section~\ref{abc}, is successful not only in reproducing photometric galaxy properties and redshift distributions for the surveys it has been designed for, but also at reproducing spectroscopic galaxy properties for a completely independent survey as SDSS. This is impressive, as the very good agreement can be seen in both the populations of red and blue galaxies when looking at the first two principal components. Differences arise when looking at the higher order components, mostly in the blue galaxy population. In Figure~\ref{fig: pca_red}, we see good to medium agreement up to the fifth principal component for the red galaxy sample. Individual features are not clearly visible, as this analysis is conducted in observed-frame. However, redshift distribution broadened absorption lines are visible, such as the G-band and H$\beta$ absorption. This whole analysis could as well be conducted in rest-frame. However, one of the purposes of this work is to provide a working framework to compare real and simulated galaxies in the absence of computed spectroscopic redshifts, as it could happen in the early phases of a spectroscopic redshift survey like the upcoming DESI.

A good agreement, but less pronounced than for red galaxies, can be seen in the blue galaxy population from the third principal component. The last three principal components have been rearranged in their order and on the mixing matrix, as the value of their variances was almost indistinguishable. This can be due to the objects statistics (4189 blue galaxy versus 6525 red galaxies), and therefore to the noise becoming more dominant with respect to the galaxy signal. More importantly, the individual features of blue galaxies, i.e, bright emission lines coming from gas physics and kinematics, are more difficult to simulate. This is due to the fact that templates like those used in this work are based on gas with constant density, metallicity and radiation fields, but galaxies are known to be a superposition of different states, which are extremely difficult to be properly taken into account, as described for example in \citep{bolatto1999, rollig2006, olsen2018}, and references therein. However, the first two principal components, which show the overall shape of the mean observed-frame spectrum, show very good agreement, proving our ability to simulate the blue galaxy population in a statistical sense. This is further proven is Figure~\ref{fig: coefficients}, which shows the distribution of eigencoefficients $ \mathrm{a}_i$ and $ \mathrm{b}_i$  of Equations~\ref{specdata} and~\ref{specsim}. It is possible to evaluate the relative contribution of each eigenspectrum to the observed spectrum by calculating the respective eigencoefficients, which are the scalar products of the eigenspectra with their normalized spectra. For both the red and the blue samples, the distributions show very good agreement in their mean and standard deviation. This is a substantial improvement from our previous analysis, which used textbook values for the luminosity function parameters, as the comparison with Figure 9 in [F18] clearly indicates.

\subsection{Galaxy Population Spectroscopic Properties}\label{secspecprop}

\begin{figure}	
	\centering
	\begin{subfigure}[t]{0.495\linewidth}
		\centering
		\includegraphics[width=1\linewidth, valign=t]{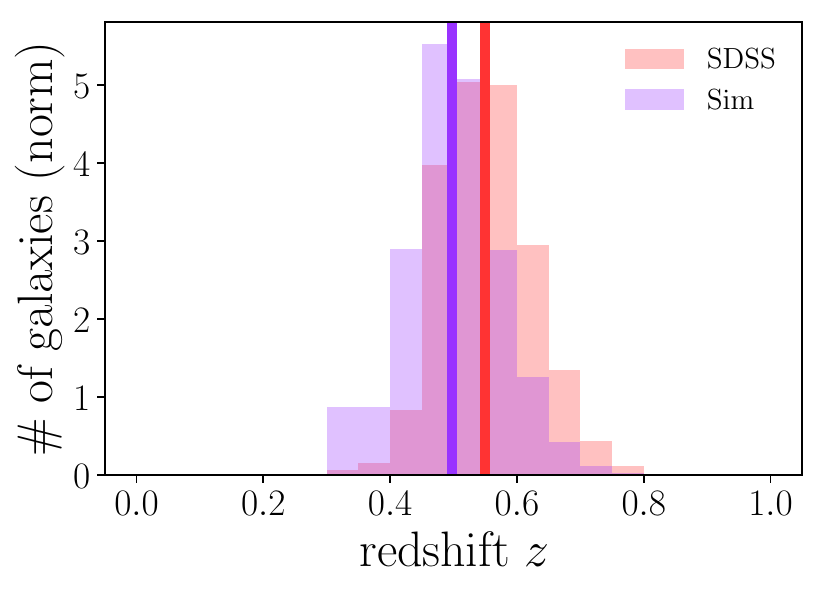}
	\end{subfigure}
	\begin{subfigure}[t]{0.495\linewidth}
		\centering
		\includegraphics[width=1\linewidth,valign=t]{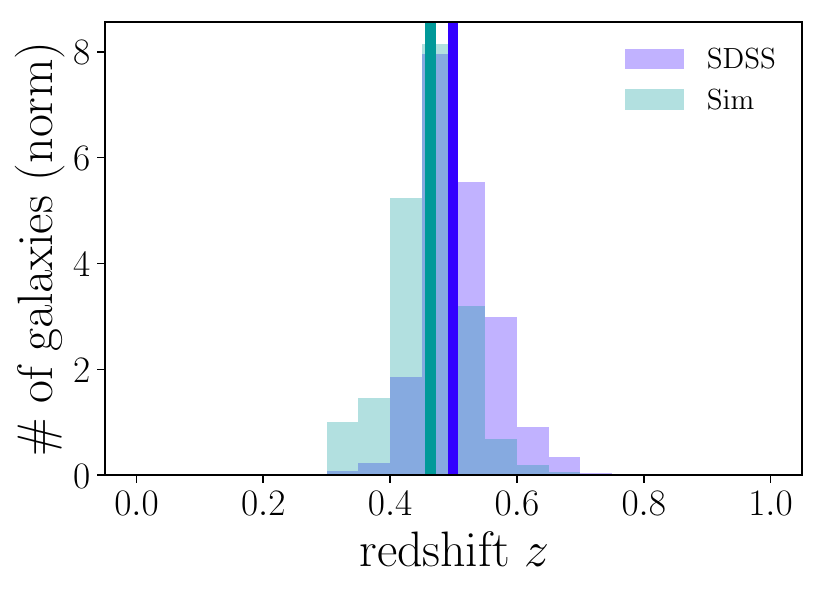}

	\end{subfigure}
	\caption{Spectroscopic redshift distribution (n$(z)$) for red (left panel), and blue (right panel) galaxies. Vertical lines represent the medians of their respective distribution of matching color. Red galaxies show a difference in their medians of $\Delta z_{\mathrm{red}}=0.052$, while blue galaxies have $\Delta z_{\mathrm{blue}}=0.035$.}
	\label{fig: reddist}
\end{figure}

We also measure individual galaxy properties and compare those. As a first step, we compare the redshift distributions n$(z)$ for red and blue galaxies. BOSS spectroscopic redshifts have been downloaded alongside with their spectra. Our simulated spectra come from our luminosity functions, as described in our model Section. Therefore, the two redshift definitions might show a small disagreement as they have not been measured in the same way. This must be taken into account when comparing them. Figure~\ref{fig: reddist} shows the distributions for red (left hand side), and blue (right hand side), galaxies. Vertical lines of the same color of their respective distributions indicate the position of their medians. The differences in the medians result in $\Delta z_{\mathrm{red}}=0.052$ for red galaxies, and $\Delta z_{\mathrm{blue}}=0.035$ for blue galaxies. These small differences in the redshift distributions can be also used to test the performances of the model presented in \citep{tortorelli2020}, and can be confronted to their Figure 13.

\begin{figure}	
	\centering
	\begin{subfigure}[t]{0.495\linewidth}
		\centering
		\includegraphics[width=1\linewidth, valign=t]{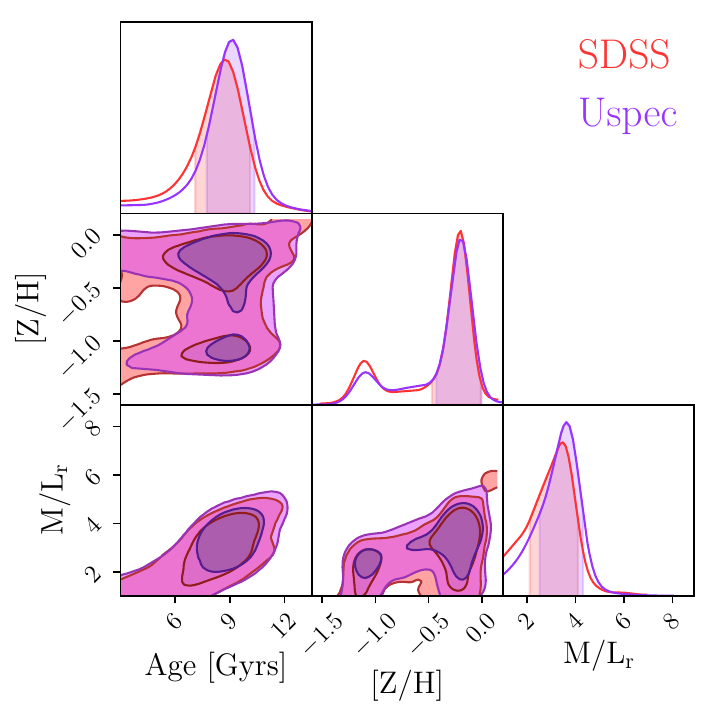}
	\end{subfigure}
	\begin{subfigure}[t]{0.495\linewidth}
		\centering
		\includegraphics[width=1\linewidth,valign=t]{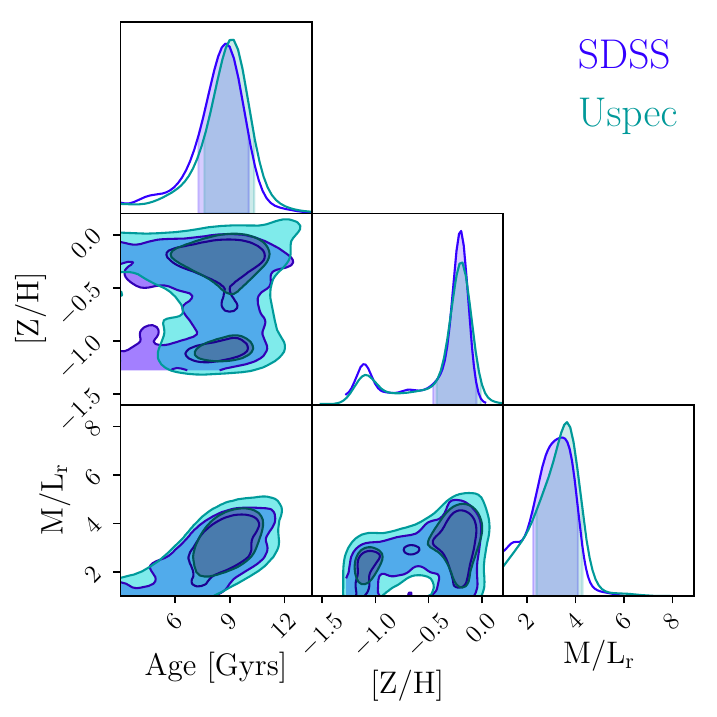}

	\end{subfigure}
	\caption{Stellar population properties comparison for red (left panel) and blue (right panel) between real and simulated galaxies. We show stellar ages (in Gyrs), total metallicity ([Z/H]), and Mass to Light ratios in the $r-$band (M/L$_r$).}
	\label{fig: gal_prop}
\end{figure}

As explained in Section~\ref{full}, we also measure stellar population properties for both red and blue galaxies, on data and simulations, and we compare the property distributions. In Figure~\ref{fig: gal_prop}, we show the comparison between stellar ages, stellar metallicities, and mass-to-light ratios in the $r-$band, for our real and simulated spectra. Stellar properties appear to be very similar for red and blue galaxies. This is due to the fact that CMASS Sparse is a sub-sample of CMASS. The similarity between the two samples is also clearly visible in Figure~\ref{fig: color_cuts}. 

On the left hand side of Figure~\ref{fig: gal_prop}, red galaxy stellar population properties are shown. Stellar ages appear to be centered at about 9-10 Gyrs, as expected for galaxies at these redshifts \citep{fagioli2016}. Real and simulated spectra show very good overlap. We find such high values for stellar ages also for blue galaxies not only because CMASS Sparse is a subsample of CMASS, but also because during the full spectral fitting we masked regions where emission lines coming from gas were dominant, avoiding in such way to be dominated by this effect when deriving stellar properties. The stellar ages derived here can be also compared to those derived through SED fitting in \citep{maraston2013}, with the caveats that different methodologies and choice of templates may lead to different results when deriving stellar population parameters. However, the values we report here are in the same range of those derived in that study.

An excellent agreement can be seen also for stellar metallicities, for both showing a double population peaked at values of slightly sub-solar metallicities, and very low (sub-solar) metallicities. This is in agreement with studies of CMASS galaxies, as stated for example in \citep{maraston2013}, where red galaxies never reach solar metallicity values. 

A good overlap is also shown for the mass-to-light ($r-$band) ratio, with simulated galaxies showing a small shift towards lower values. The same considerations can be drawn for the blue galaxies population, with a slightly more marked disagreement for the mass-to-light ratio values. 

The agreement between our real and simulated galaxies, already visible from the PCA statistical analysis, is therefore also confirmed by looking at individual stellar galaxy population properties, further strengthening our considerations about the goodness of the model presented in \citep{tortorelli2020}, and our technique in simulating galaxy spectra.

\section{Discussion on Forward Modeling}

\begin{figure}	
	\centering

		\includegraphics[width=157mm]{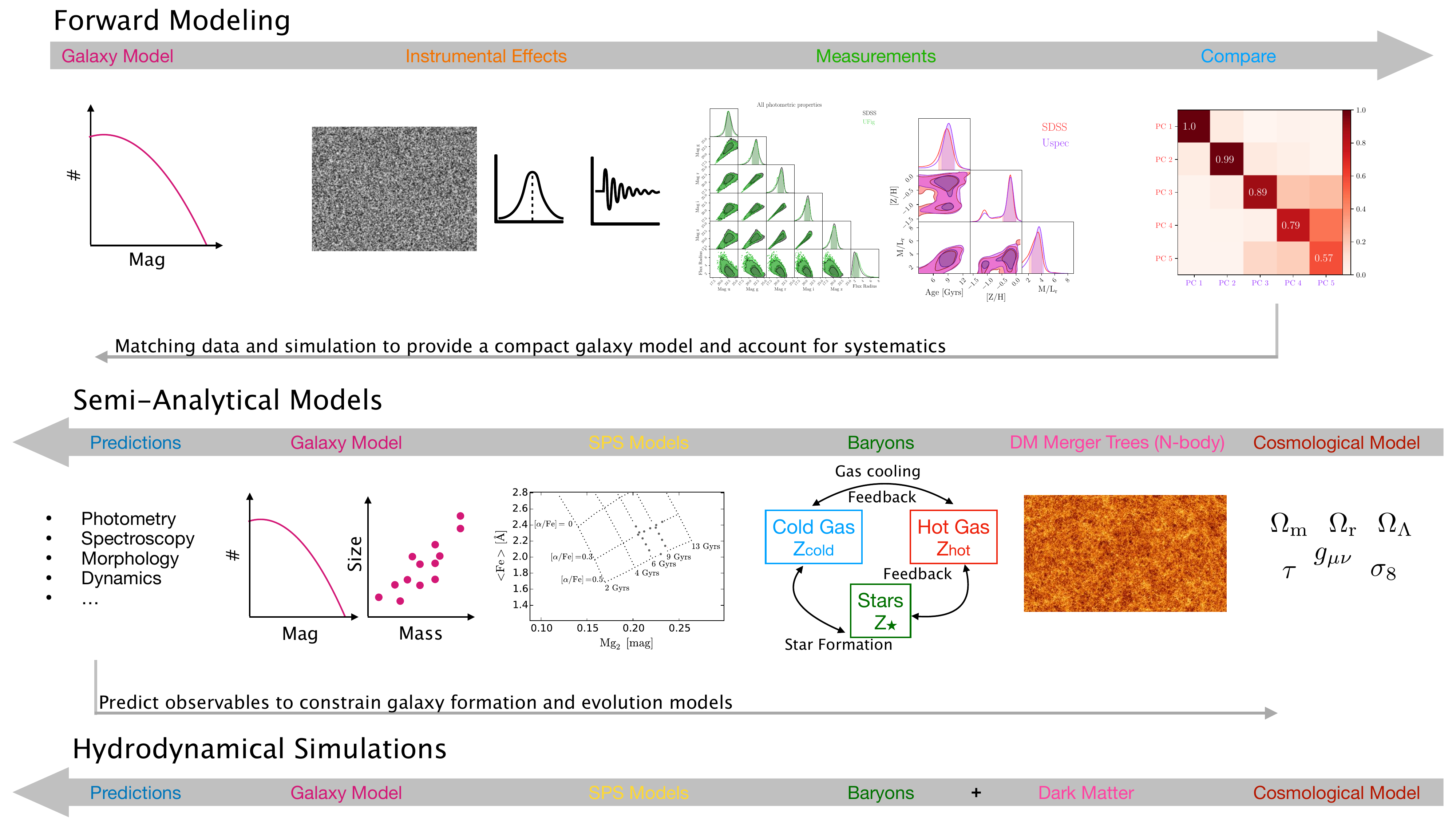}

	\caption{Synopsis of the different approaches used to model the galaxy population. Our forward modeling approach, as well as 
semi-analytical models and hydrodynamical simulations are shown. Forward modeling (top row) starts from a galaxy model, then includes 
instrumental parameters, and measurements of galaxy properties on data and simulations which are compared statistically. If the agreement between data and simulations is not sufficient, the process is repeated using modified input parameters. The process takes about a few hours on a supercomputer in our case. The second and third rows show the semi-analytical and the hydro-simulations approaches. In semi-analytical models, a cosmological model is first chosen. Then, using N-body simulations, the merger trees of dark matter halos of different masses are traced. At this point, baryons are inserted in the model together with the effects they give rise to such as hot gas cooling, cold gas conversion into stars, and feedback. This is translated into galaxy properties thorough Stellar Population Synthesis (SPS) codes. A complex, physically motivated galaxy formation and evolution model can then be developed, and different predictions about galaxy and gas properties, metallicity, temperature, initial mass function, dynamics, and morphology can be made. In hydrodynamical simulations, a cosmology is also chosen as a first step. The density fields of dark and baryonic matter are modeled. The evolution of these density fields is followed by solving numerically gravitational and hydrodynamical equations using prescriptions for baryonic physics. As well as semi-analytical models, they provide a wide-range of predictions on galaxy formation and evolution.}
	\label{fig:flowchart}
\end{figure}

As the analysis presented above shows promising results, we discuss here our forward modeling technique and how it compares to other approaches.

Our analysis makes use of a simpler model for the galaxy population than the models used in semi-analytical models such as \cite{white1978, white1991, baugh2006, benson2010a, benson2010b, monaco2014, hou2019} or in hydrodynamical simulations such as \cite{vogelsberger2014, schaye2015}. As illustrated in Figure~\ref{fig:flowchart} and described in Section~\ref{model}, we indeed use a galaxy population model as an input. This includes prescriptions for the luminosity functions of red and blue galaxies, as well as a simple model for galaxy sizes and SED distribution functions, as described in \citep{tortorelli2020}. In \cite{tortorelli2020}, it was shown that these prescriptions are successful in reproducing current observations taken from \citep{giallongo2005, ilbert2006, zucca2009, loveday2012, cool2012, fritz2014, beare2015}. A question that arises is how this simple model performs for further properties which are notoriously difficult to reproduce, such as stellar ages, stellar metallicities, and mass-to-light ratios. In our approach, these properties are not inputs in the model, but they are measured after the fact in simulations and data. Therefore, we do not necessarily measure intrinsic stellar properties, but we show how our simulated sample is statistically similar to a real BOSS sample of spectra after accounting for all selection biases and systematics, for images and spectra.

Results presented in Figure~\ref{fig: gal_prop} (see also Figure~\ref{fig: magnitudes_cuts}) show properties measured consistently on data and simulations, after careful modeling of the noise properties of images and spectra, and accounting for selection effects coming from possibly inconsistent definitions of magnitudes. These results show that our method provides a successful treatment of selection biases. Stellar population properties in particular are especially prone to biases, as results obtained with different techniques (full spectral fitting vs. Lick System, for example) can be different within the same data set \citep{fagioli2016}. Even when using the Lick System of spectral lines only, the use of different lines can lead to different results in absolute numbers, as shown in \cite{fagioli2016}, however preserving consistent relative results (for example, in \cite{fagioli2016}, smaller galaxies are consistently older than bigger galaxies, for stellar masses $10.5 < \log\mathrm{M}_*/\mathrm{M}_\odot < 11$, but the absolute ages change with different approaches, as shown in Table 3 in \citep{fagioli2016}). 

Figure~\ref{fig:flowchart} shows a comparison between our forward modeling approach, and approaches based on semi-analytical models and hydrodynamical equations. This flowchart highlights the goals of the different methods. Forward modeling (top row) starts from a galaxy model, adding various instrumental parameters, then measures quantities on data and simulations, and then compares the two resulting measurement sets. If the level of agreement is not satisfactory, the process is repeated using a modified set of input parameters until an agreement is found using control loops. As the software tools involved are fast, this whole procedure can be executed within a few hours on a supercomputer. If the level of agreement cannot be reached by modifying both instrumental and galaxy population parameters, then the model is complexified and another set of control loops is started.

The second and third rows of Figure~\ref{fig:flowchart} illustrate the semi-analytical and  hydrodynamic-simulation approaches. In semi-analytical models, a cosmological model is first chosen. Then, using N-body simulations, the merger trees of dark matter halos of different masses are traced. At this point, baryons are inserted in the model together with the effects they give rise to such as hot gas cooling, cold gas conversion into stars, and feedback. A complex, physically motivated galaxy formation and evolution model can then be developed, and different predictions about galaxy and gas properties, metallicity, temperature, initial mass function, dynamics, and morphology can be made. In hydrodynamical simulations, a cosmology is also chosen as a first step. The density fields of dark and baryonic matter are modeled by solving numerically gravitational and hydrodynamical equations. As well as semi-analytical models, hydrodynamic-simulations provide a wide-range of predictions on galaxy formation and evolution. These predictions go beyond the goals of our forward modeling scheme. Also, it should be taken into account that when differences arise in the comparison between the predictions of such models to real data, it is often difficult to attribute them to specific physical processes, or to noise, systematics or selection effects. Running these simulations is also a much slower process that in our forward modeling approach, especially when referring to hydrodynamical simulations.

In \citep{gonzalez2009}, the authors show the predictions of the \texttt{GALFORM} \citep{cole2000} semi-analytical galaxy formation model for the luminosities, morphologies, colours and scale-lengths of local galaxies. In order to compare their results to SDSS data, the authors needed to convert the standard \texttt{GALFORM} outputs into SDSS-like properties, like for example Petrosian magnitudes. Independently on how well one can make such a conversion, the effects of noise on data are important, especially when considering magnitudes in different bands. We showed in \citep{fagioli2018} that even when carefully modeling the noise in each band, it is difficult to obtain an agreement between real and simulated magnitudes. This is because modeling the effect of the noise in each band analytically is challenging. Such an effect can cause colours to be incorrectly estimated, and therefore galaxies associated to wrong categories in the colour-colour space. For example, our findings indicate that a large part of the differences in galaxy properties such as magnitudes and colours we saw in our previous galaxy model (see Figures 1 and 6 in [F18], but also abundance ratios in Appendix C) can be ascribed to measurements systematics and selection biases [F18]. This aspect can be explored by using, as a starting point, a simple galaxy model which can be complexified as needed, and which can be compared to other approaches. In this work, we applied the same procedure to compute real and simulated magnitudes at the image level. Therefore, with our current method we can safely assume that our residual differences between data and simulations are coming from the model rather than other effects.

Forward modeling is therefore meant to be a complementary approach to semi-analytical models and hydrodynamical simulations. The two latter approaches provide a more detailed modeling of the galaxy population, but are computationally slower than our approach which provides a detailed treatment of instrumental and selection effects. Our approach thus provides a compact, simple galaxy population model which is corrected for systematic effects and can be a useful input for comparison with the other approaches. 

\section{Conclusions}\label{conclusions}

In this paper, we present a method to forward model a spectroscopic galaxy survey, using a target selection based on a simulated wide field imaging survey. This method can be used to simulate both red and blue galaxies. 

We start from a galaxy population model which provides us with luminosity function parameters. Luminosity functions are different for red and blue galaxies. From those distributions, we draw magnitudes, redshifts and spectral coefficients, which constitute the basis for the construction of our simulated images and spectra. We then simulate a SDSS-like imaging survey with \texttt{U\textsc{fig}}, using 157,000 images coming from the DR14 sample. We pre-process and analyse the images to derive all the parameters needed to simulate SDSS-like images. This whole analysis and simulation only take up to $\sim4$h with 1,000 cores of a supercomputer. We simulate SDSS-like images with our fast image simulator, \texttt{U\textsc{fig}}. We then apply selection cuts to both data and simulations. This is a particularly important step, as this allows us to avoid biases coming from different definitions of magnitudes or star-galaxy separation. We apply cuts with using our own measured magnitudes, corrected for PSF effects, and CLASS\_STAR parameters coming from \texttt{SE\textsc{xtractor}}. 

We then retrieve SDSS/BOSS DR14 spectra. We simulate their analogs with our own built-in software devoted to that, called  \texttt{U\textsc{spec}}. We repeat the same procedure for what we call the red galaxy sample, constructed with emulating the CMASS sample cuts, and what we call the blue galaxy sample, which we obtained by following the CMASS Sparse sample cuts, a bluer subsample of the parent CMASS.

We use two main procedures to compare our populations of real and simulated galaxies. By using PCA, we compare the main statistical properties of the different samples. The first two principal components, which encode most of the information regarding the galaxy population shape of the spectrum, show very good agreement for both red and blue galaxies. The same can be said for the distribution of all the eigencoefficients, which are centered to each other and have the same width. Differences arise when looking at higher order principal components, which are more sensitive to the features of individual galaxies. These differences appear to be more pronounced for blue galaxies, as expected by the notorious difficulty to simulate their individual emission features. Also, differences in the noise properties might be relevant for higher order principal components, especially for blue galaxies, which have a lower number statistics than red galaxies.

We then compare the redshift distribution n($z$) of red and blue galaxies, finding good agreement overall but small differences in their medians. Also, we measure individual stellar population properties with \texttt{pPXF}, namely stellar ages, stellar metallicities and stellar masses, for both red and blue galaxies, finding very good agreement for both, and values in agreements with previous CMASS studies. 

These results are a clear indication that our galaxy population model not only works for the wide-field imaging surveys it has been designed for, but also on completely independent spectroscopic data. 

Our ability to reproduce realistic galaxy spectra for both red and blue galaxies offers very good prospects for future spectroscopic redshift surveys. Not only the inclusion of galaxy clustering properties can be used for cosmology applications, but the realistic stellar population properties we are able to reproduce can be used to connect clustering to studies of galaxy formation and evolution.

\acknowledgments
We acknowledge support by Swiss National Science Foundation (SNF) grant 200021\_169130. AR is grateful for the hospitality of KIPAC at Stanford University/SLAC where part of his contribution was made. MF would like to thank Elisabetta Ghisu and Tomasz Kacprzak for discussions on PCA, and Pascale Berner and Beatrice Moser for helpful discussions on the manuscript.  
This work has made use of data from the European Space Agency (ESA) mission
{\it Gaia} (\url{https://www.cosmos.esa.int/gaia}), processed by the {\it Gaia}
Data Processing and Analysis Consortium (DPAC,
\url{https://www.cosmos.esa.int/web/gaia/dpac/consortium}). Funding for the DPAC
has been provided by national institutions, in particular the institutions
participating in the {\it Gaia} Multilateral Agreement.
This research has made use of the VizieR catalogue access tool, CDS, Strasbourg, France (DOI: 10.26093/cds/vizier). The original description of the VizieR service was published in \citep{ochsenbein2000}. This research made use of \textsf{IP\textsc{ython}}, \textsf{\textsc{NumPy}},  \textsf{\textsc{SciPy}}, and \textsf{\textsc{Matplotlib}}.


\bibliographystyle{JHEP.bst}
\bibliography{paper_fagioli_bluegal.bbl}



\end{document}